\documentstyle[preprint,aps,floats]{revtex}
\tightenlines
\input psfig.tex 
\begin{document}
\def\ba{\begin{eqnarray}}
\def\ea{\end{eqnarray}}
\def\be{\begin{equation}}
\def\ee{\end{equation}}
\def\({\left(}
\def\){\right)}
\def\lagrange {{\cal L}}
\def\del {\nabla}
\def\d {\partial}
\def\Tr{{\rm Tr}}
\def\half{{1\over 2}}
\def\fourth{{1\over 8}}
\def\bibi{\bibitem}
\def\S{{\cal S}}
\def\H{{\cal H}}
\def\xx{\mbox{\boldmath $x$}}
\newcommand{\phpr} {\phi_0^{\prime}}
\newcommand{\gam}{\gamma_{ij}}
\newcommand{\sqgam}{\sqrt{\gamma}}
\newcommand{\dph}{\delta\phi}
\newcommand{\om} {\Omega}
\newcommand{\dom}{\delta^{(3)}\left(\Omega\right)}
\newcommand{\rar}{\rightarrow}
\newcommand{\Rar}{\Rightarrow}
\newcommand{\labeq}[1] {\label{eq:#1}}
\newcommand{\eqn}[1] {(\ref{eq:#1})}
\newcommand{\labfig}[1] {\label{fig:#1}}
\newcommand{\fig}[1] {\ref{fig:#1}}
\def\gsim{ \lower .75ex \hbox{$\sim$} \llap{\raise .27ex \hbox{$>$}} }
\def\lsim{ \lower .75ex \hbox{$\sim$} \llap{\raise .27ex \hbox{$<$}} }
\newcommand\bigdot[1] {\stackrel{\mbox{{\huge .}}}{#1}}
\newcommand\bigddot[1] {\stackrel{\mbox{{\huge ..}}}{#1}}
%\twocolumn[\hsize\textwidth\columnwidth\hsize\csname @twocolumnfalse\endcsname

\title{Gravitational Waves in Open de Sitter Space}

\author{S.W. Hawking\thanks{S.W.Hawking@damtp.cam.ac.uk},
Thomas Hertog\thanks{Aspirant FWO-Vlaanderen;
%\hspace{-.05in}
email:T.Hertog@damtp.cam.ac.uk} and
Neil Turok\thanks{email:N.G.Turok@damtp.cam.ac.uk} \\ $\ $ \\
DAMTP \\
Centre for Mathematical Sciences \\
Wilberforce Road, Cambridge, CB3 0WA, UK.}
\date{\today}
\maketitle

\begin{abstract}

We compute the spectrum of primordial gravitational wave perturbations in open 
de Sitter spacetime. The background spacetime is taken to be the 
continuation of an O(5) symmetric instanton saddle point of the
Euclidean no boundary path integral.
The two-point tensor fluctuations are 
computed directly from the Euclidean path 
integral. The Euclidean correlator is then analytically continued into the
Lorentzian region where it describes the quantum mechanical vacuum
fluctuations of the graviton field.
Unlike the results of earlier work, the correlator is shown to be unique 
and well behaved in the infrared. We show that the
infrared divergence found in previous calculations is due
to the contribution of a discrete
gauge mode inadvertently included in the spectrum.

\end{abstract}
\vskip .2in

\section{Introduction}
 
One appeal of inflationary cosmology is its mechanism for the origin of
cosmological perturbations. The de Sitter phase of exponentially-rapid 
expansion quickly redshifts away any local perturbations, leaving behind 
only the quantummechanical vacuum fluctuations in the various fields. 
During inflation, these perturbations are stretched to macroscopic length 
scales and subsequently amplified, to later seed the growth of the large scale
structures in the present-day universe.
A particularly clean example of this effect are the gravitational wave
perturbations of the spacetime itself.
These tensor perturbations contribute to the cosmic microwave background 
anisotropy via the Sachs-Wolfe effect. They may 
potentially provide an observational discriminant between different theories 
of open (or closed) inflation because their long-wavelength modes
strongly depend on the boundary conditions at the instanton
that describes the beginning of the inflationary universe \cite{Hertog}.

Although the tensor spectrum has been successfully computed in realistic 
$O(3,1)$ invariant models for an open inflationary universe \cite{Hertog}, 
the problem of 
calculating the primordial gravitational waves in perfect open de Sitter
spacetime has remained a paradox for some time. 
The previous literature claims that the spectrum of gravitational waves 
in perfect de Sitter space is infrared divergent 
for all physically well-motivated initial quantum states of an
eternally inflating universe
\cite{allencald,Tanaka,Mottola}.
Breaking the 
$O(4,1)$ invariance of de Sitter space by going to a realistic 
inflationary model
introduces a potential barrier for the tensor fluctuation modes,
and it has been argued that the bubble wall acts to
regularise the divergent spectrum in perfect de Sitter space
\cite{Tanaka}.

Previous calculations of the gravitational wave spectrum \cite{allencald,Tanaka}
in open de Sitter space are based on a mode-by-mode analysis.
One has a prescription for the vacuum state of the graviton that is
imposed on every mode separately, on some Cauchy surface for the de Sitter 
spacetime. Then one propagates each mode into the open universe region.
In this paper we instead compute the two-point tensor
correlator in real space. In doing so, we have obtained an infrared finite
tensor spectrum. The difference in the two approaches is related
to the non-uniqueness of the mode decomposition in an open universe,
as we shall explain.

As an aside, we mention in this context that also fluctuations of a massless 
minimally coupled scalar field in de Sitter space do not break
$O(4,1)$. In some prior literature (see e.g. \cite{Bruce}) it is shown that
there is no de Sitter invariant propagator for such a scalar field.
However, the scalar field is not itself an observable since the action 
depends only on its derivative, and there is a symmetry 
$\phi \rightarrow \phi +$ constant. In fact, correlators of space or time
derivatives of $\phi$ are de Sitter invariant, and since these are the only 
physical correlators in the theory, de Sitter invariance is unbroken.

We implement the Hartle--Hawking no boundary proposal
\cite{Hartle} in our work by 'rounding off' open de Sitter space on a compact
Euclidean instanton, namely a round four sphere.
The fluctuations are computed in the Euclidean region
directly from the Euclidean path integral, 
to first order in $\bar h$ around the instanton saddle point.
The Euclidean two-point correlator is analytically 
continued into the Lorentzian region where it describes the quantum 
mechanical vacuum fluctuations of the graviton field
in the state described by the no boundary 
proposal initial conditions.
There is no ambiguity in the choice of initial
conditions because the Euclidean correlator is unique.

\section{Tensor Fluctuations about Cosmological Instantons}

In quantum cosmology the basic object is the wavefunctional $\Psi [h_{ij},
\phi ]$, the amplitude for
a three-geometry with metric
$h_{ij}$ and field configuration $\phi$. It is formally 
given by a path integral
\ba
\label{uni}
\Psi \left[h_{ij},\phi \right] \sim
\int^{h_{ij},\phi}\left[{\cal D} g\right] \left[{\cal D}\phi \right] e^{iS[ g,\phi ] }.
\ea

Following Hartle and Hawking 
\cite{Hartle} the lower limit of
the path integral is defined by 
continuing to Euclidean time and integrating over all compact 
Riemannian metrics $g$ and
field configurations $\phi$. If one can find a saddle point of (\ref{uni}),
namely a classical solution satisfying the Euclidean no boundary condition,
one can in principle at least compute the path integral 
as a perturbative expansion to any desired power in  $\hbar$.

In this paper we wish to compute the two-point 
tensor fluctuation correlator in open de Sitter spacetime,
\begin{equation}
ds^2  =  -dt^2 + \sinh^2 (t)\left(d\chi^2 +\sinh^2 (\chi)
d\Omega^2_2) \right).
\end{equation}
Open de Sitter space may be obtained by analytic continuation of an
O(5) invariant instanton, describing the beginning of a
semi-eternally inflating universe. 
The analytic continuation is given by setting
$t=-i\sigma$ and the radial coordinate $\chi=i\Omega$, where
$\Omega$ is the polar angle on the three sphere (see \cite{Gratton}).
The instanton obtained in this way is a solution of the Euclidean 
equations of motion with the maximal symmetry allowed in four dimensions. 
It takes the form of a round four sphere
with line element 
$ds^2 =d\sigma^2 +\sin^2 (\sigma) d\Omega^2_3 $, where $d\Omega^2_3$ is 
the line element on $S^3$.
It is useful to introduce a
conformal spatial coordinate $X$ defined by
$\int_{\sigma}^{\pi/2} \frac{d\sigma'}{\sin\sigma'}$,
%$dX=d\sigma/\sin(\sigma)$, 
so that the line element takes the form 
\begin{eqnarray}
ds^2 & = & \cosh^{-2} X \left( dX^2+ d\Omega_3^2\right).
\end{eqnarray}
On the four sphere $X$ then ranges from $-\infty$ to $+\infty$.

The principles of our method to calculate cosmological perturbations
are described in detail in
\cite{Hertog,Gratton}.
The instanton solution
provides the classical background with respect to which the quantum 
fluctuations are defined. 
In the Euclidean region the exponent $iS$ in the path integral
becomes $-S_E=-(S_0+S_2)$, 
where $S_E$ is the Euclidean action, $S_0$ is the 
instanton action 
and $S_2$ the action for fluctuations. We keep the latter
only to second order. The path integral for
the two-point tensor fluctuation about a particular instanton background
is then given by 
\ba\label{wave}
\langle t_{ij}(x)t_{i'j'}(x')\rangle =
{ \int \left[{\cal D} \delta g\right]\left[{\cal D} \delta \phi\right]
e^{-S_2}t_{ij}(x)t_{i'j'}(x') \over  \int \left[{\cal D} \delta g\right]
\left[{\cal D} \delta \phi\right] 
e^{-S_2}}.
\ea
To first order in $\bar h$ the quantum fluctuations are
specified by a Gaussian integral.
The Euclidean action
determines 
the allowed perturbation modes because divergent modes are suppressed
in the path integral.
The Euclidean two-point tensor correlator is 
then analytically 
continued into the Lorentzian region where it describes the quantum 
mechanical vacuum fluctuations of the graviton field
in the state described by the no boundary 
proposal initial conditions.

To find the perturbed action $S_2$ that enters in the path integral 
(\ref{wave}), we 
write the perturbed line element in open de Sitter space as
\begin{equation}\label{line}
ds^2  =  \sinh^{-2} (\tau) \left(- ( 1+2A)d\tau^2 + S_{i}dx^{i} d\tau +
(\gamma_{ij} + h_{ij}) dx^{i}dx^{j} \right),
\end{equation}
where the fields $A$, $S_{i}$ and $h_{ij}$
are small perturbations.
Because we are interested in the gravitational wave spectrum in the open slicing of
de Sitter space, we will only retain $O(3,1)$ invariance in our calculation.

The quantities $S_{i}$ and $h_{ij}$ may be uniquely decomposed
as follows \cite{Stewart},
\begin{eqnarray}\label{dec}
h_{ij} & = & \frac{1}{3} h \gamma_{ij} + 2 \left(
\nabla _{i} \nabla_{j}  - \frac{\gamma_{ij}}{3} \Delta_{3}\right) E
+ 2 F_{(i \vert j)} + t_{ij},\nonumber\\
S_{i} & = & B_{\vert i} + V_{i}.
\end{eqnarray}
Here $\Delta_{3}$ is the Laplacian on $S^3$ and $\vert j$ the 
covariant derivative on the three-sphere. With respect to
reparametrisations of the three-sphere, $h$, $B$ and
$E$ are scalars, $V_{i}$ and $F_{i}$ are divergenceless vectors
and $t_{ij}$ is a transverse traceless symmetric tensor, describing the
gravitational waves.
Because gauge transformations are scalar or vector, the
perturbations $t_{ij}$ are automatically gauge invariant.

It is important to note that the gauge invariance of $t_{ij}$ follows
from the uniqueness of the above decomposition. This is only true
however for bounded (asymptotically decaying) perturbations \cite{Stewart}. 
If one does not impose suitable
asymptotic conditions on the fields, a degeneracy appears between scalar and
tensor perturbations that introduces a discrete gauge mode 
in the tensor spectrum, which plays a crucial role in the
divergent behaviour of the correlator. We come back to this point in 
Section V.

We now substitute the decomposition (\ref{dec}) into the 
Lorentzian action for gravity plus a cosmological constant,
\begin{equation}\label{action}
S = \frac{1}{2\kappa} \int d^4 x \sqrt{-g}\left( R 
- 2\Lambda\right) -
\frac{1}{\kappa}\int d^3 x \sqrt{\gamma} K,
\end{equation}
The scalar, vector and tensor quantities decouple.
Keeping all terms to second order, we continue the perturbed Lorentzian
action to the Euclidean region. The scalar and vector fluctuations are pure
gauge in perfect de Sitter space.
The tensor perturbations $t_{ij}$ 
yield the following well-known positive Euclidean action
\cite{Mukhanov}:
\begin{equation}\label{action2}
S_{2} =  
\frac{1}{8\kappa}
\int d^4 x \frac{\sqrt{\gamma}}{\cosh^2 X}  \left(
t'^{ij}t'_{ij} + t^{ij\vert k}t_{ij\vert k} + 2 t^{ij}t_{ij} \right).
\end{equation}
Here prime denotes differentiation with respect to the conformal coordinate 
$X$.  After performing the rescaling
$\tilde t_{ij} = \frac{t_{ij}}{\cosh X}$ and integrating by parts 
we obtain
\begin{equation}\label{act}
S_{2} = 
\frac{1}{8\kappa}
\int d^4 x \sqrt{\gamma}
\tilde t_{ij} \left(
 \hat K+3 -\Delta_{3} \right)\tilde t^{ij}
+{1\over 8\kappa} 
\left[\int d^3x \sqrt{\gamma}  \tilde 
t_{ij}\tilde t^{ij} \tanh(X)\right],
\end{equation}
where the Schr\"{o}dinger operator 
\begin{equation}\label{pot}
\hat K = -\frac{d^2}{dX^2}  - \frac{2}{\cosh^2 (X)} 
\equiv  -\frac{d^2}{dX^2} +U(X)
.\end{equation}

Because the fluctuations are specified by a Gaussian integral, we can
solve the path integral (\ref{wave}) 
by looking for the Green function of the operator in its exponent.
The potential $U(X)$ for the fluctuation modes
is well known to be perfectly reflectionless.
However, changing its shape slightly would introduce some reflection
which becomes increasingly significant at small momenta.
Such a change corresponds to breaking the 
$O(5)$ invariance of Euclidean 
de Sitter space and is exactly what happens in the O(4) invariant
Hawking--Turok \cite{Hawking} and Coleman--De Luccia \cite{Coleman}
instantons that describe the beginning of realistic
open inflationary universes.
This difference between both classes of instantons
has profound implications for the
tensor perturbations about them, especially for their
long-wavelength regime \cite{Hertog}. 
The operator $\hat K$ has in all three cases
a positive continuum starting at eigenvalue
$p^2 = 0$, as well as a single bound state $\tilde t_{ij} = b(X)q_{ij}(\Omega)$
at $p=i$ which turns out to be a trivial gauge mode.
%\hfill\break 
%\begin{figure}
%\centerline{\psfig{file=potential.ps,width=2.5in}}
%\caption{} 
%\end{figure}

\section{The Euclidean Green Function}

To evaluate the path integral (\ref{wave}), we first 
look for the Green function 
$G_{E}^{\ iji'j'} (X,X',\Omega,\Omega')$
of the operator in (\ref{act}). The Euclidean fluctuation 
correlator (\ref{wave})
will then be given by $\cosh(X)\cosh(X')G_{E}^{\ iji'j'}$.
The Euclidean Green function satisfies
\begin{equation}\label{green}
\frac{1}{4\kappa} \left(
\hat K +3 -\Delta_{3} \right)G^{\ ij}_{E\ i'j'}
(X,X',\Omega , \Omega ')
=\delta (X-X') \gamma^{-{1\over 2}} \delta^{ij}_{\ \ i'j'}(\Omega - \Omega ').
\end{equation}
If we think of the scalar product as defined by integration over $S^3$
and summation over tensor indices, then 
the right hand side is the normalised projection operator onto
transverse traceless tensors on $S^3$.

The Green function $G_{E\ i'j'}^{\ ij}$ can only be a function of 
the geodesic distance
$\mu (\Omega , \Omega')$ if it is to be invariant under isometries of the
three-sphere. 
This suggests that
\begin{equation}\label{ans}
G^{\ ij}_{E\ i'j'}(\mu,X,X') = 
4\kappa \sum_{p=3i}^{+i\infty} G_{p}(X,X') W^{\ ij}_{(p)\ i'j'}(\mu ),
\end{equation}
where $W^{\ ij}_{(p)\ i'j'}(\mu)$ is a bitensor that is invariant under the 
isometry group O(4).
It equals the sum (\ref{bit}) of the normalised rank-two tensor
eigenmodes with eigenvalue $\lambda_p = p^2 +3$ of the Laplacian on $S^3$.
Note that the indices $i,j$ lie in the tangent space over the point $\Omega$
while the indices $i',j'$ lie in the tangent space over the point $\Omega'$.
On $S^3$ we have
\begin{equation}
\Delta_{3} W^{\ ij}_{(p)\ i'j'}(\mu) = \lambda_{p}W^{\ ij}_{(p)\ i'j'}(\mu).
\end{equation}
The motivation for the unusual labelling of the eigenvalues of the Laplacian
is that, as demonstrated in the Appendix, in terms of the label $p$ the 
bitensor on $S^3$ has precisely the same formal expression as the
corresponding bitensor on $H^3$. It is precisely this property that 
will enable us in Section IV to
continue the Green function from the 
Euclidean instanton into open de Sitter space without decomposing it
in Fourier modes.
The relation between the bitensors on $S^3$ and $H^3$
together with some useful formulae and properties of maximally
symmetric bitensors
are given in Appendix A. 

Since the tensor eigenmodes of the Laplacian on $S^3$ form a complete
basis, we can also write
\begin{equation}\label{delta}
\gamma^{-{1\over 2}}
\delta^{ij}_{\ \ i'j'}(\Omega - \Omega ') = \sum_{p=3i}^{+i\infty} 
W^{\ ij}_{(p)\ i'j'}(\mu(\Omega,\Omega')).
\end{equation}
Hence by substituting our ansatz (\ref{ans}) for the Green function
into (\ref{green}) we obtain an
equation for the 
X-dependent part of the Green function,
\begin{equation}\label{xgreen}
\left(\hat K -p^2\right) G_{p}(X,X') = \delta (X-X').
\end{equation}
The solution to equation (\ref{xgreen}) is 
\ba
 G_{p}(X,X') &=&
{1\over \Delta_p} \left[\Psi_{p}^{r}(X) \Psi_{p}^{l}(X')\Theta(X-X')+
\Psi_{p}^{l}(X) \Psi_{p}^{r}(X')\Theta(X'-X)\right].
\ea 
$\Psi_{p}^{l}(X)$ is the solution to the Schr\"{o}dinger 
equation that tends to $e^{-ipX}$ as $X \rightarrow -\infty$, and 
$\Psi_{p}^{r}(X)$ is the solution going as $e^{ipX}$ 
as $X \rightarrow +\infty$.  The factor $\Delta_p$ is 
the Wronskian of the two solutions.
Since the potential is reflectionless on
the round four sphere the left- and right-moving waves do not 
mix and they
equal the Jost functions $g_{\pm p}(X)$ with nice analytic properties.
The solutions may be found explicitely 
and are given by
\begin{equation}\label{explic}
\left\{
\begin{array}{lll}
\Psi_{p}^{r}(X) &=&
(\tanh X - ip)e^{ipX}\\
\Psi_{p}^{l}(X) &=&
(\tanh X + ip)e^{-ipX}\\
\end{array}
\right.
\end{equation}
and their Wronskian $\Delta_p =-2ip(1+p^2)$, independent of $X$.
The zero of the Wronskian at $p=i$
corresponds to the bound state mentioned above. 
Taking $X>X'$, we obtain the Euclidean Green function as a discrete sum
\begin{equation}\label{disc}
G_{E}^{\ iji'j'}(\mu,X,X') =4\kappa
\sum_{p=3i}^{i\infty} \frac{i}{2p}\frac{\Psi_{p}^{r}(X)\Psi_{p}^{l}(X')}
{(1+p^2)}
W^{\ iji'j'}_{(p)}(\mu).
\end{equation}

Before proceeding, let us demonstrate that the
Euclidean Green function is regular at the poles of the four sphere.
This is a nontrivial check because the coordinates $\sigma$ and $X$ are 
singular there, and the rescaling becomes divergent too.
In the large $X, X'$ limit, (\ref{disc}) becomes
\begin{equation}\label{sum}
G_{E}^{\ iji'j'}(\mu,X,X') = 
2 \kappa \sum_{n=3}^{\infty}\frac{1}{n}e^{-n(X - X')}W^{\ iji'j'}_{(in)}(\mu)
\end{equation}
For $n \geq 3$ the Gaussian hypergeometric functions  $F(3+n,3-n,7/2,z)$ that
constitute the bitensor $W^{\ iji'j'}_{(n)}$ have a series
expansion that terminates, and they 
essentially reduce
to Gegenbauer polynomials $C^{(3)}_{n-3}(1 - 2z)$.
Using then the identity \cite{Erdelyi}
\begin{equation}
\sum_{l=0}^{\infty} C^{\nu}_{l}(x)q^{l} = \left( 1 -2xq + q^2 \right)^{-\nu}
\end{equation}
with $q = e^{-(X- X')}$, one easily sees that the sum
(\ref{sum}) indeed converges. 

We have the Euclidean Green function defined as an infinite sum (\ref{disc}).
However, the eigenspace of the Laplacian on $H^3$ suggests that the Lorentzian
Green function is most naturally expressed as an integral over real $p$.
To do so we must extend the summand into the upper half $p$-plane. 
We have already defined the wavefunctions $\Psi_{p}(X)$ as analytic functions
for all complex $p$ but we need
to extend the
bitensor as well. When the Green function is expressed as a discrete sum,
it involves the bitensor $W^{\ iji'j'}_{(p)}(\mu )$ evaluated at $p=ni$ with
$n$ integral. At these values of $p$, the bitensor is regular at both
coincident and opposite points on $S^3$, that is at $\mu=0$ and $\mu=\pi$.
However, if we extend $p$ 
into the complex plane we lose regularity at 
$\mu=0$, essentially  because the bitensor obeys the differential equation 
(\ref{green}) with
a delta function source at $\mu =0$.
Similarly
we must maintain regularity at $\mu=\pi$, since there is no delta
function source there. 
The condition of regularity at $\pi$ imposed by the differential
equation for the Green function is sufficient
to uniquely specify the analytic continuation of $W^{\ iji'j'}_{(in)}(\mu )$
into the complex $p$-plane.
The continuation is described in the Appendix,
and the extended bitensor $W^{\ iji'j'}_{(p)}(\mu )$ is defined by
equations (\ref{bitensor}) and (\ref{app1}).

Now we are able to  
write the sum in (\ref{disc}) 
as an integral along a contour ${\cal C}_1$ encircling
the points $p=3i, 4i, ... Ni$, where $N$ tends to infinity. 
For $X>X'$ we have
\begin{eqnarray}
G_{E}^{\ iji'j'}(\mu,X,X') & = & \kappa\int_{{\cal {C}}_1} \frac{dp}{p 
\sinh p\pi}\frac{\Psi_p^{r}(X) \Psi_p^{l}(X')}{(1+p^2)} 
W^{\ iji'j'}_{(p)}(\mu).
\label{discint}
\end{eqnarray}
To see that (\ref{discint}) is equivalent to the sum (\ref{disc}) 
introduce $1=\cosh p \pi /\cosh p \pi$ into the integral.
Then note that 
$\coth p\pi$ has 
residue $\pi^{-1}$ at every integer
multiple of $i$. Finally, use (\ref{relat}) to
rewrite $W^{\ iji'j'}_{(p)}(\mu)$ in the form regular 
at $\mu=0$ used in (\ref{disc}). The factor of $\cosh p \pi$ from (\ref{relat})
cancels that in the integrand. 

We now distort the contour for the $p$ integral to run 
along the real $p$ axis (Figure 1). At large imaginary $p$
the integrand decays exponentially 
and the contribution vanishes in the limit of large $N$. 
However as we deform the contour towards the real axis 
we encounter two poles in the $\sinh p\pi$ factor, the 
latter at $p=i$ becoming a
double pole due to the simple zero of the Wronskian. 
For the $p=2i$ pole, it follows from
the normalisation of the tensor harmonics 
that $W_{(2i)}^{iji'j'}=0$. 
Indirectly, this is a consequence of 
the fact that spin-2 perturbations do not have a monopole or dipole component.
At $p=i$ we have a double pole, but
although the relevant Schr\"{o}dinger operator
possesses a bound state, it
does not generate a `super-curvature mode'.
Instead the relevant mode is a time-independent 
shift in the metric perturbation
which may be gauged away
\cite{Hertog,Tanaka}.
We conclude that up to a term involving a pure gauge mode, we can deform the 
contour ${\cal {C}}_1$ into the contour shown in Figure 1.
For the moment, since the integrand involves a factor $p \sinh p\pi$ which 
has a double pole at $p=0$, we leave the contour avoiding the origin 
on a small semicircle in the upper half $p$-plane. 

\begin{figure}
\centerline{\psfig{file=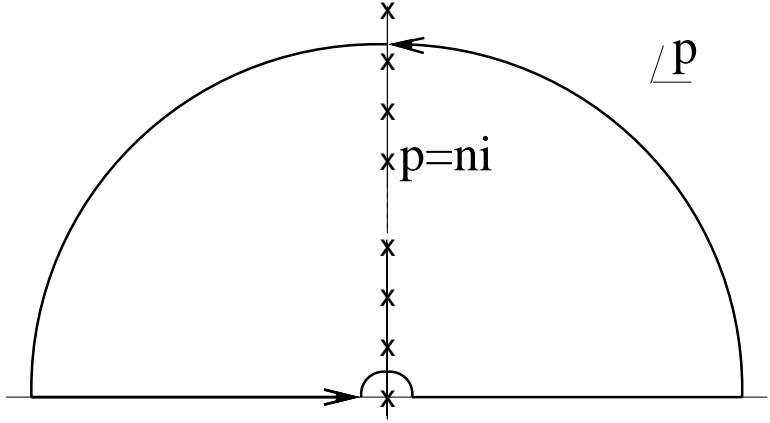,width=4in}}
\caption{Contour for the Euclidean Correlator.}
\end{figure}

Finally, in order to deal with the pole at $p=0$, we re-express the
integrand in (\ref{discint}) as a sum of its
$p$-symmetric and $p$-antisymmetric parts. Denoting the integrand
by $I_{p}$ we then have
\begin{equation}
G_{E}^{iji'j'} =
\frac{1}{2}\int dp (I_{p} + I_{-p}) +\frac{1}{2}\int dp (I_{p} - I_{-p}), 
\end{equation}
where the integral is taken from $p=-\infty$ to $\infty$ along a path
avoiding the origin above. But
$\int dp I_{-p}$ along this contour is equal to the integral of $I_{p}$ 
taken along a contour avoiding
the origin below.
The second term is therefore equal to the integral of $I_{p}$ along
a contour around the origin. Hence we have
\begin{equation}
\frac{1}{2}\int dp (I_{p} - I_{-p}) =
-\pi i {\bf Res }(I_{p};p=0).
\end{equation}

We defer a detailed discussion of this term to Section V, because its
interpretation is clearer in the Lorentzian region. Hence for the time being we
just keep it, but it will turn out that it represents a non-physical 
contribution to the graviton propagator.

In the $p$-symmetric part of the correlator, 
we can leave the integrand as a sum of $I_{p}$ and $I_{-p}$.
We henceforth 
denote the path from $-\infty$ to $+\infty$ avoiding the origin above
by ${\cal R}$. This shall turn out to be a regularised version of
the integral over the real axis.
Our final result for the Euclidean Green function
then reads
\begin{eqnarray}\label{ecor}
G^{E}_{\ iji'j'}(\mu,X,X') & = & \frac{\kappa}{2}\int_{{\cal R}} {dp \over p
\sinh p \pi} 
\frac{W_{\ iji'j'}^{(p)}(\mu)}{(1+p^2)}\left(
\Psi_{p}(X) \Psi_{-p}(X')+\Psi_{-p}(X) \Psi_p(X') \right)\nonumber\\
 & & -\pi i{\bf Res }(I_{p};p=0).
\end{eqnarray}

\section{Two-Point  Tensor Correlator in Open de Sitter Space}

The analytic continuation into open de Sitter space 
is given by setting $\sigma=it$ and the polar angle
$\Omega=-i\chi$.
Without loss of generality we may take one of the two 
points, say $\Omega'$ to be at the north pole of the three-sphere. Then $\mu 
= \Omega$, and $\mu$ continues to $-i\chi$. 
We then obtain the correlator in open de Sitter space where one point 
has been chosen as the origin of the radial coordinate $\chi$.
%The background line element of the Lorentzian region is
%\ba
%ds^2 = -dt^2 + \sinh^2(t) \left( d\chi^2 + \sinh^2 \chi d\Omega_2^2 \right).
%\ea
The conformal coordinate $X$ continues to conformal time $\tau$ 
as $X=-\tau - \frac{i\pi}{2}$ (see \cite{Gratton}).
%\ba
%X \equiv \int_{it}^{\pi/2} \frac{d\sigma}{\sin(\sigma)} = - \tau - \frac{i
%\pi}{2}
%\label{contx}
%\ea
%where the conformal time $\tau$ is defined in \cite{Gratton}.
%\ba
%\ \tau \equiv \lim_{\epsilon \rightarrow 0} \left( \int_{\epsilon}^{\pi/2}
%\frac{d\sigma}{\sin(\sigma )} - \int_{\epsilon}^{t}\frac{dt'}{\sinh(t')}
%\right).
%\label{taudef}
%\ea

Hence the analytic continuation of the Euclidean mode functions is
given by 
\ba
\Psi_{p}^{r}(X) \rightarrow -e^{p \pi \over 2} \Psi^L_{p}(\tau)
\quad \mathrm{and} \quad 
\Psi_{p}^{l}(X) \rightarrow -e^{-p \pi \over 2} \Psi^L_{-p}(\tau)
\labeq{jostcont}
\ea
where the Lorentzian mode functions are
\begin{equation}\label{fun}
\Psi_p^L(\tau) 
=(\coth \tau + ip)e^{- ip\tau}.
\end{equation}
They are
solutions to the Lorentzian perturbation equation
$\hat{K} \Psi_p^L(\tau)
= p^2 \Psi_p^L(\tau)$.

In order to perform the substitution $\mu = -i\chi$, where $\chi$ is the
comoving separation on $H^3$,
we use the explicit formula given in the appendix
for the bitensor regular at $\mu=\pi$.
The continued bitensor $W_{iji'j'}^{(p)}(\chi)$ is defined by the
equations (\ref{app1}), (\ref{app3a}) and (\ref{app3}).
%We obtain the following $p$-integral for the Lorentzian Green function:
%\begin{eqnarray}
%\label{eucor}
% G_{L}^{iji'j'}(\chi,X,X') & = & 
%\kappa \int_{C}\frac{dp}{p\sinh p\pi}
%\frac{\Psi_{p}^{L}(\tau)\Psi_{-p}^{L}(\tau')}{(1+p^2)}W^{\ iji'j'}_{(p)}
%(\chi).\end{eqnarray}
It can be seen from (\ref{app3}) that it
involves terms which behave as $e^{\pm p(i\chi + \pi)}$.
One must extract the $e^{p\pi}$-factors in order for the bitensor to 
correspond to the usual sum of rank-two tensor harmonics on the real $p$-axis.
To do so we use the following general identity.
For $\tau'-\tau>0$, we have (up to the $p=i$ gauge mode)
\ba
\int_{\cal C} {dp \over p} 
{\Psi_{p}^{L}(\tau) \Psi_{-p}^{L}(\tau') \over (1+p^2)} e^{i p\chi} F(p)
= 0,
\label{intid}
\end{eqnarray}
where $F(p)$ are the $p$-dependent coefficients
occurring in the final (Lorentzian) form of
the bitensor given in (\ref{app4}). This identity follows from the analyticity 
of the integrand. By inserting
$1= \sinh p\pi/ \sinh p\pi$ under the integral, it is clear that
the integral (\ref{intid}) with  a factor $e^{p\pi}/\sinh p \pi$ 
inserted equals that with a factor $e^{-p\pi}/\sinh p \pi$ inserted. 
The resulting identity allows us to replace the factors $e^{+p(i\chi+\pi)}$
in the bitensor by $e^{p(i\chi- \pi)}$, and vice versa in the analog integral
of $I_{-p}$ closed in the lower half $p$-plane. 

For the tensor correlator 
we also need to restore the factor $ia^{-1}(\tau)$ to $t_{ij}$.
It is convenient to define the eigenmodes
$\Phi_{p}^{L}(\tau) = \Psi_{p}^{L}(\tau)/a(\tau)$. The extra minus sign 
hereby introduced is cancelled
by a change in sign of the normalisation factor $Q_p$ of the
bitensor, which then becomes
$+(p^2 +4)/(30\pi^2)$. 
This corresponds to requiring the spacelike metric to have postive signature.
We finally obtain the Lorentzian tensor Feynman (time-ordered)
correlator, for $\tau'-\tau >0$, 
\begin{eqnarray}
\label{lorcor}
\langle  t_{ij}(x),t_{i'j'}(x')\rangle & = & 
\frac{\kappa}{2}         
\int_{R}\frac{dp}{p \sinh p\pi}\frac{W_{\ iji'j'}^{L(p)}(\chi)}{(1+p^2)}
\left(e^{-p\pi}\Phi_{p}^{L}(\tau )\Phi_{-p}^{L}(\tau ')
+e^{p\pi}\Phi_{-p}^{L}(\tau )\Phi_{p}^{L}(\tau ')\right)\nonumber\\
 &  & -\pi i {\bf Res } (I^{L}_{p};p=0),
\end{eqnarray}
where the Lorentzian
bitensor $W_{iji'j'}^{L(p)}$ is defined in the Appendix, equations
(\ref{bitensor}) and (\ref{app4}).

In this section, we concentrate on the first term in
(\ref{ecor}), the integral over $p$, and
ignore for the moment the second, discrete term.
We first extract the symmetrised part,
$\langle \{ t_{ij}(x), t_{i'j'}(x') \} \rangle$, which is just
the real part of the Feynman correlator.
The imaginary part involves an integrand which is analytic 
for $p \rightarrow 0$:
\begin{eqnarray}\label{rewriii}
\langle  t_{ij}(x),t_{i'j'}(x')\rangle &  = & 
{\kappa\over 2} \int_{\cal R} {dp \over p(1+p^2)}
W_{\ iji'j'}^{L(p)}(\chi)
 {\rm coth}  p\pi [
\Phi_p^L(\tau) \Phi_{-p}^L(\tau') +
\Phi_{-p}^L(\tau) \Phi_{p}^L(\tau')]\nonumber\\
& & 
- 2\kappa \int_{0}^{\infty} dp 
\frac{W_{\ iji'j'}^{L(p)}(\chi)}{(1+p^2)}
{\cal I } \left[ \frac{1}{p}\Phi^{L}_{p}(\tau)\Phi^{L}_{-p}(\tau')
\right].
\end{eqnarray}

It is straightforward to
see that if we apply the Lorentzian version of the perturbation operator
$\hat K$ to (\ref{rewriii}) with an appropriate heaviside function of
$\tau-\tau'$, the imaginary term will produce the Wronskian of 
$\Phi_{-p}^L(\tau)$
and $\Phi_{p}^L(\tau)$, which is proportional to $ip$,
times $\delta(\tau-\tau')$. Then the integral over $p$ produces a spatial
delta function. From this one sees that our
Feynman correlator obeys the correct
second order partial differential equation, with a delta function source.
The delta function source term in (\ref{green}) 
goes from being real in the Euclidean region
to imaginary in the Lorentzian region because 
the factor $\sqrt{g}$ continues to  $i\sqrt{-g}$. 

The integral in (\ref{lorcor}) diverges as $p^{-2}$ for
$p \rightarrow 0$, in contrast with realistic
models for inflationary universes where a reflection term in (\ref{rewriii})
regularises the spectrum \cite{Hertog}.
However, as we immediately show, even in perfect de Sitter
space the integral over $p$ is perfectly finite.
We rewrite the symmetrised correlator as an integral over real
$0 \leq p \leq \infty$ as follows.
Because the integrand in (\ref{rewriii}) is even in $p$, we have
\begin{eqnarray}
\langle \{ t_{ij}(x),t_{i'j'}(x')\}\rangle & = &
2\kappa 
\int_{\epsilon}^\infty \frac{dp}{\pi p^2}\frac{p\pi \coth p\pi}{(1+p^2)}
\Re \left[\Phi_{p}^{L}(\tau )\Phi_{-p}^{L}(\tau ')\right]
W_{\ iji'j'}^{L(p)}(\chi)\nonumber \\
& &
-\frac{2\kappa}{\pi \epsilon}
\Phi_{0}^{L}(\tau)\Phi_{0}^{L}(\tau ')W_{\ iji'j'}^{L(0)}(\chi)
+O(\epsilon),
\end{eqnarray}
the second term being the contribution from the small semicircle around 
$p=0$.
Both terms may be combined under one integral. The resulting integrand is
\emph{analytic} as $p \rightarrow 0$ and one can safely take the limit
$\epsilon \rightarrow 0$.
The symmetrised correlator is then given by
\begin{center}
$\langle \{ t_{ij}(x),t_{i'j'}(x')\}\rangle = $
\end{center}
\begin{equation} \label{lorcorf}
2\kappa 
\int_0^\infty \frac{dp}{\pi p^2}\left(\frac{p\pi \coth p\pi}{(1+p^2)}
\Re \left[\Phi_{p}^{L}(\tau )\Phi_{-p}^{L}(\tau ')\right]
W_{\ iji'j'}^{L(p)}(\chi)
-\Phi_{0}^{L}(\tau)\Phi_{0}^{L}(\tau ')
W_{\ iji'j'}^{L(0)}(\chi)\right),
\end{equation}
where the Lorentzian
bitensor $W_{iji'j'}^{L(p)}$ is defined in the Appendix, equations
(\ref{bitensor}) and (\ref{app4}).
In this integral it may be written as
\begin{equation}
W^{L(p)}_{iji'j'}(\chi) =
\sum_{{\cal P}lm} q_{ij}^{(p){\cal P}lm}(\Omega ) q^{(p){\cal P}lm}_{i'j'}
(\Omega ')^{*}.
\end{equation}
The functions 
$ q_{ij}^{(p){\cal P}lm}(\Omega ) $ are the rank-two tensor eigenmodes 
with eigenvalues 
\newline
$\lambda_{p}=-(p^2+3)$ of the Laplacian on $H^3$.
Here ${\cal P}={e,o}$ labels the parity, and $l$ and $m$ are the usual
quantum numbers on the two-sphere.
At large $p$, the coefficient functions $w_{j}^{(p)}$ of the bitensor
(see Appendix A) behave like $p\sin p\chi$. Hence
the above integral converges at large $p$, 
for both timelike and spacelike 
separations. Furthermore, 
the correlations asymptotically decay for large separation of
the two points.

Equation (\ref{lorcor}), with the first term given by
(\ref{lorcorf}) is our final result for the two-point
tensor correlator in open de Sitter space, with Euclidean no boundary
initial conditions. 
Contracting the propagator with the harmonics
$q^{i'j'}_{(p)elm}$ and integrating over the three sphere reveals
that the second term leaves the spectrum completely unchanged apart from
cancelling the (divergent) contribution from the $p^2 =0$ divergence in the
first term.
We defer a detailed discussion of this result to the next section, 
in which we will also clarify the difficulties of the
previous work on the graviton propagator in open de Sitter 
spacetime \cite{allencald,Tanaka,Mottola}.

As an illustration let us compute the Sachs-Wolfe
integral \cite{Sachs} and show that all the multipole moments 
are finite. 
The contribution of gravitational 
waves to the CMB anisotropy in perfect de Sitter
space is given by
\begin{eqnarray}\label{temp}
\frac{\delta T_{SW}^{\ }}{T}(\theta,\phi)
& = & -\frac{1}{2}\int_{0}^{\tau_{0}}d\tau
t_{\chi \chi,\tau}^{\ }(\tau,\chi,\theta,\phi)\vert_{\chi=
\tau_0 - \tau},
\end{eqnarray}
where $\tau_{0}$ is the observing time.
The temperature anisotropy on the sky
is characterised by the two-point angular correlation function
$C(\gamma)$, where $\gamma$ is the angle
between two points located on the celestial sphere.
It is customary to expand the correlation function in terms of
Legendre polynomials as
\begin{equation}\label{ang}
C(\gamma) =\left\langle \frac{\delta T}{T}(0)\frac{\delta T}{T}(\gamma)
\right\rangle =
\sum_{l=2}^{\infty} \frac{2l+1}{4\pi} C_{l} P_{l}(\cos \gamma)
.\end{equation}
Hence, inserting the Sachs-Wolfe integral into (\ref{ang}) and substituting
(\ref{lorcorf}) for the two-point fluctuation correlator yields the
multipole moments
\begin{eqnarray}\label{multi}
C_{l}  & = &
\frac{\kappa}{2}  \int_{0}^{+\infty}dp
\int_{0}^{\tau_{0}}d\tau
\int_{0}^{\tau_{0}}d\tau'
\left(\frac{\coth p\pi}{p(1+p^2 )} 
\Re \left[\dot \Phi^{L}_{p}(\tau)\dot \Phi^{L}_{p}(\tau')\right]
Q^{pl}_{\chi\chi}
Q^{pl}_{\chi'\chi'} \right. \nonumber\\
& & \qquad \qquad \qquad \qquad \qquad \qquad \qquad
\qquad \qquad \qquad \left.
 - \dot \Phi^{L}_{0}(\tau) \dot \Phi^{L}_{0}(\tau')
Q^{0l}_{\chi\chi}
Q^{0l}_{\chi'\chi'} \right).
\end{eqnarray}
In this expression we have written the normalised tensor 
harmonics  $q_{\chi\chi}^{(p)elm}(\chi,\theta,\phi)$ as
$Q^{pl}_{\chi\chi}(\chi)Y_{lm}(\theta, \phi)$, where
\begin{equation}
Q^{pl}_{\chi\chi}(\chi) = \frac{N_{l}(p)}{p^2(p^2+1)} 
(\sinh \chi)^{l-2} \left(\frac{-1}{\sinh \chi}\frac{d}{d\chi}\right)^{l+1}
(\cos p\chi)
\end{equation}
and
\begin{equation}
N_{l}(p) =\left[\frac{(l-1)l(l+1)(l+2)}{\pi \prod_{j=2}^{l}(j^2+p^2)}
\right]^{1/2}.
\end{equation}
It can readily be seen that the multipole moments are finite.
With the aid of the explicit expressions and the wavefunctions (\ref{fun}) 
they can be numerically computed.

\section{Conclusions}

We have computed the spectrum of primordial 
gravitational waves predicted in open de Sitter
space, according to Euclidean no boundary initial conditions. The Euclidean 
path integral unambiguously specifies the tensor fluctuations with no 
additional assumptions. The real space Euclidean correlator has been 
analytically continued into the Lorentzian region without
Fourier decomposing it, and
we obtained an infrared finite two-point tensor 
correlator in open de Sitter space,
contrary to previous results in the literature
\cite{allencald,Tanaka,Mottola}.

Let us now elaborate on the second, regularising term in the symmetrised
correlator (\ref{lorcorf}) and the discrete $p=0$ contribution to the Feynman
correlator given from the last term in (24).
Not surprisingly, they have a similar interpretation.
Their angular part $W_{\ iji'j'}^{L(0)}(\chi)$ 
is equal to the sum of the tensor harmonics with eigenvalue 
$\lambda_{p}(p=0)= -3$ of the Laplacian on $H^3$.
It has been known that a degeneracy appears between $p^2 =0$ tensor 
modes and $p_{s}^2 =-4$ scalar harmonics \cite{Tanaka}. More 
specifically, one has $q_{ij}^{e(0)lm}=
\left( \nabla_{i}\nabla_{j} - \frac{1}{3} \gamma_{ij} \nabla^{2}\right)
q^{(2i)lm}$ 
where $q^{(2i)lm}=  P_{(2i)lm}Y_{lm}$. 
The discrete $p^2=0$ tensor harmonics are the only transverse traceless
tensor perturbations that can be constructed from a scalar quantity.
But as a consequence of this, they are sensitive
to scalar gauge transformations.
Consider now the coordinate transformation
$\xi^{\alpha}=(0,\epsilon\Phi^{L}_{0}(\tau)\nabla^{i}q^{(2i)lm})$. 
Under this transformation the transverse 
traceless part of the metric perturbation $h_{ij}$ in the perturbed
line element (\ref{line}) changes exactly by 
$\epsilon t_{ij}^{(0)lm}=\epsilon\Phi^{L}_{0}(\tau)q^{(0)lm}_{ij}$. 
Using the transverse-traceless properties of $t_{ij}$ it is
easily seen that the action for tensor fluctuations
is invariant under such transformations.
Hence this tensor eigenmode is non-physical and can be gauged away.
Note that since the functional form of $\xi$ is
completely fixed this corresponds to a global transformation,
analogous to the transformation  $\phi \rightarrow \phi +$ constant 
for a massless field. To compute the Green function for
a massless field one has to project out this homogeneous mode,
and it is necessary to do the same here.
One should therefore disregard the contribution from the discrete term in (24)
to the Lorentzian correlator. This was actually also done in our computation
of the tensor fluctuation spectrum about $O(4)$ instantons \cite{Hertog},
although in that case not because the mode was pure gauge, but 
because it couples to the inflaton field, and is not represented by a simple 
action
of the form (8). If a scalar field is present, the mode is most simply
treated as a part of the scalar perturbations, as was done in \cite{Gratton}.

In our result (\ref{lorcorf}) 
for the symmetrised correlator, the discrete gauge mode 
is set to zero because the second term cancels exactly the contribution from 
the $p^2 =0$ mode implicitly contained in the continuous spectrum.
This automatic cancellation does not happen in the conventional
mode-by-mode analysis where, if one chooses the most degenerate
continuous representation of the isometry group $O(3,1)$ of the 
hyperboloid $H^3$, corresponding to the range $p \in [0,\infty )$, one
obtains a divergent correlator.

It is clear that the underlying reason for these subtleties has to do with the
different nature of tensor harmonics on compact and non-compact spaces.
Hence, we could have expected the generation of the two discrete gauge modes
simply from the analytic continuation of the completeness relation (14) of 
the harmonics on $S^3$. Apart from the sum of the complete set of modes that 
constitute the delta function on $H^3$, one obtains
also three extra terms $W^{iji'j'}_{(2i)}(\mu)$, $W^{iji'j'}_{(i)}(\mu)$
and $W^{iji'j'}_{(0)}(\mu)$. The first term is zero, and the
remaining two terms should respectively be viewed as sums
of vector - and scalar harmonics. On the other hand, the fact that
the scalar/tensor degeneracy appears precisely at the lower bound of the 
continuous spectrum is a peculiar feature of three dimensions. In the
analogous computation in four dimensions for instance \cite{reall},
this degeneracy happens at $p^2 =-1/4$ and consequently, there is no
regularising term in the correlator.

There is yet another way in which the exclusion of the degenerate modes
from the perturbation spectrum can be interpreted.
Remember that in non-compact spacetimes
the decomposition (\ref{dec}) is only uniquely defined for
bounded perturbations. Hence, 
the only way there can appear a degeneracy between
the different types of fluctuations 
is for the degenerate modes to be unbounded.
Indeed, on the three-hyperboloid the scalar $p^2=-4$ modes
describe divergent fluctuations because the scalar spherical harmonics
$q^{(2i)lm}$ grow exponentially with distance.
The action of the above tensor operator renders only the $q_{\chi j}^{(0)lm}$
components of $q_{ij}^{(0)lm}$ finite at infinity.
The remaining components still diverge as $\sim e^{\chi}$ and correspond to
exponentially growing fluctuations at large distances\footnote{The 
confusion arises because, due to the form of 
the metric inverse, scalar invariants are finite at infinity, e.g.
$q_{ij}q^{ij} \sim e^{-2\chi}$. This also explains why the coefficient
functions $w_{j}^{(0)}(\chi)$ in the bitensor $W^{L(0)}_{iji'j'}$ 
asymptotically decay.}.
Since in cosmological perturbation theory one assumes the
perturbation $h_{ij}$ to be small, one must expand correlators
in bounded harmonics.

We want to emphasize that the regularity of the two-point tensor correlator
does not depend on the Euclidean methods used in our work. 
One could have equally well computed the correlator on closed Cauchy surfaces
for the de Sitter space where the subtleties encountered here do not arise,
assuming the standard conformal vacuum for that slicing. One would then
analytically continue the result to the open slicing.
On the other hand, the Euclidean no boundary principle is an appealing
prescription which avoids the arbitrary choice of vacuum otherwise needed.
The path integral effectively
defines its own initial conditions, yielding a
unique and infrared finite Green function in the Lorentzian region. 
The initial quantum state of
the perturbation modes, defined by the no boundary path integral, 
corresponds to the conformal vacuum in the Lorentzian spacetime. 
This is in many ways the most natural state in de Sitter space, but
the regularity of the
graviton propagator is independent of this choice.
The most important technical advantage of our method is that
we deal throughout directly with the real space correlator, which
makes the derivation independent of the gauge ambiguities involved in
the mode decomposition.

Finally, let us conclude by comparing the gravitational wave spectrum in perfect open
de Sitter spacetime with the spectrum in realistic open
inflationary universes.
In both the Hawking--Turok and the Coleman--De Luccia model for open
inflation there is an extra reflection term in the correlator
because O(5) symmetry is broken on the instanton \cite{Hertog}.
This term gives rise to long-wavelength
bubble wall fluctuations in the Lorentzian region.
At first sight, the wall fluctuations seem 
to regularise the spectrum.
However, adding and subtracting the second term in (\ref{lorcorf})
to the two-point tensor correlator in the $O(4)$ models
(eq. (34) in \cite{Hertog}) and
comparing that with our result (\ref{lorcorf}) reveals
that the wall fluctuations actually appear as an extra 
long-wavelength continuum contribution 
\emph{on top of} the spectrum in perfect de Sitter space.
Hence in both the Hawking--Turok and Coleman--De Luccia model
there is an enhancement of the
fluctuations compared to the perturbations
in perfect de 
Sitter space. But the singularity in Hawking--Turok instantons 
suppresses the wall fluctuations because
it enforces Dirichlet boundary conditions on the perturbation modes
\cite{Hertog}. Hence we expect the spectrum in perfect de Sitter
space to be quite similar to the spectrum predicted by singular instantons.
On the other hand, Coleman--De Luccia models typically predict large
wall fluctuations, yielding a very different CMB anisotropy spectrum on
large angular scales.
The tensor fluctuation spectrum therefore potentially
provides an observaional discriminant between different theories of
open inflation \cite{gratthertog}.

\bigskip
\centerline{\bf Acknowledgements}

It is a pleasure to thank Steven Gratton and Valery Rubakov
for stimulating discussions.

\appendix            

\section{Maximally Symmetric Bitensors}

A maximally symmetric bitensor $T$ is one for which $\sigma^{*}T=0$
for any isometry $\sigma$ of the maximally symmetric manifold.
Any maximally symmetric bitensor may be expanded in terms of a complete set of
'fundamental' maximally symmetric bitensors with the correct index symmetries.
For instance
\begin{eqnarray}\label{maxi}
T_{iji'j'} &  = & 
t_1(\mu) g_{ij}^{\ }g_{i'j'}^{\ }+
t_2(\mu)\left[n_{i}^{\ }g_{ji'}^{\ }n_{j'}^{\ }+
n_{j}^{\ }g_{ii'}^{\ }n_{j'}^{\ }+ n_{i}^{\ }g_{jj'}^{\ }n_{i'}^{\ }+
n_{j}^{\ }g_{ij'}^{\ }n_{i'}^{\ }\right]\nonumber\\
& & +t_3(\mu)\left[ g_{ii'}^{\ }g_{jj'}^{\ }+g_{ji'}^{\ }g_{ij'}^{\ }
\right]+ t_4(\mu)n_{i}^{\ }n_{j}^{\ }n_{i'}^{\ }n_{j'}^{\ }
+t_5(\mu)\left[g_{ij}^{\ }n_{i'}^{\ }n_{j'}^{\ }+n_{i}^{\ }n_{j}^{\ }
g_{i'j'}^{\ }\right]
\end{eqnarray}
where the coefficient functions $t_{j}(\mu)$ depend only on the 
distance $\mu(\Omega,\Omega')$ along the shortest geodesic
from $\Omega$ to $\Omega'$.
$n_{i'}^{\ }(\Omega,
\Omega ')$ and $n_{i}^{\ }(\Omega, \Omega ')$ are
unit tangent vectors to the geodesics joining $\Omega$ and $\Omega'$ and
$g_{ij'}(\Omega, \Omega ')$ is the parallel propagator along the 
geodesic; $V^{i}g_{i}^{j'}$ is the vector at $\Omega'$ obtained by
parallel transport of $V^{i}$ along the geodesic from $\Omega$ to $\Omega'$
\cite{Jacob}.

The set of tensor eigenmodes on $S^3$ or $H^3$
forms a representation of the symmetry group of the
manifold. It follows in particular that 
their sum over the parity states  ${\cal P}=\{e,o\}$ and the
quantum numbers $l$ and $m$ on the two-sphere defines a maximally symmetric 
bitensor on $S^3$ (or $H^3$) \cite{Jacob}
\begin{equation}\label{bit}
W^{\ ij}_{(p)\ i'j'}(\mu) =
\sum_{{\cal P}lm} q^{(p)ij}_{{\cal P}lm}(\Omega ) q^{(p){\cal P}lm}_{i'j'}(\Omega ')^{*}.
\end{equation}
On $S^3$ the label $p=3i,4i,..$. It is related to the usual angular momentum 
$k$
by $p=i(k+1)$. The ranges of the other labels is then
$0 \leq l \leq k$ and $-l \leq m \leq l$.
On $H^3$ there is a continuum of eigenvalues $p \in 
[ 0, \infty)$. 
We will assume from now 
that the eigenmodes on are normalised by the condition
\begin{equation}\label{norm}
\int \sqrt{\gamma} d^{3} x q^{(p)ij}_{{\cal P}lm}
q_{{\cal P}'l'm'ij}^{(p')*} = \delta^{pp'}\delta_{{\cal P}{\cal P}'}
\delta_{ll'}\delta_{mm'}
\end{equation}

The bitensor $W^{\ ij}_{(p)\ i'j'}(\mu)$
appearing in our Green function has some additional 
properties arising from its construction in terms of the transverse and 
traceless tensor harmonics $q_{ij}^{(p){\cal P}lm}$.
The tracelessness of $W^{(p)}_{iji'j'}$
allows one to eliminate two of the coefficient functions in
(\ref{maxi}).
It
may then be written as
\begin{eqnarray}\label{bitensor}
W^{(p)}_{iji'j'}(\mu) & = & 
w^{(p)}_1\left[ g_{ij}^{\ } -3n{i}^{\ }n_{j}^{\ }\right]
\left[g_{i'j'}^{\ } -n_{i'}^{\ }n_{j'}^{\ }\right]
+ w_2^{(p)}\left[4n_{(i}^{\ }g_{j)(i'}^{\ }n_{j')}^{\ }
+4n_{i}^{\ }n_{j}^{\ }n_{i'}^{\ }n_{j'}^{\ }
\right]\cr
& & 
+w_3^{(p)}\left[ g_{ii'}^{\ }g_{jj'}^{\ }+g_{ji'}^{\ }g_{ij'}^{\ }
 -2n_{i}^{\ }g_{i'j'}^{\ }n_{j}^{\ } -2n_{i'}^{\ }g_{ij}^{\ }n_{j'}^{\ }
+6n_{i}^{\ }n_{j}^{\ }n_{i'}^{\ }n_{j'}^{\ }\right]
\end{eqnarray}
This expression is traceless on either index pair $ij$ or $i'j'$.
The requirement that the bitensor be transverse
$\nabla^{i}W_{iji'j'}^{(p)}=0$ and the
eigenvalue condition $(\Delta_{3} - \lambda_{p})W^{\ iji'j'}_{(p)}=0$
impose additional constraints on the remaining coefficient functions
$w_{j}^{(p)}(\mu)$. To solve these constraint equations it is convenient to
introduce the new variables \cite{Allen} on $S^3$ (on $H^3$, $\mu$ is replaced
by $-i\tilde \mu$)
\begin{equation}\label{bet}
\left\{
\begin{array}{lll}
\alpha(\mu) & = &  w_1^{(p)}(\mu) + 
w_3^{(p)}(\mu)\\
\beta(\mu)&  = & \frac{7}{(p^2 +9)\sin \mu}\frac{d\alpha(\mu)}{d\mu}
\end{array}
\right.
\end{equation}
In terms of a new argument $z=\cos^2(\mu/2)$ (or its continuation
on $H^3$) the transversality and eigenvalue conditions imply 
for $\alpha(z)$
\begin{equation}\label{hyper}
z(1-z)\frac{d^2\alpha(z)}{d^2z} + \left[ \frac{7}{2} -7z\right]
\frac{d\alpha(z)}{dz}=(p^2 +9)\alpha(z)
\end{equation}
and then for the coefficient functions
\ba\label{app1}
\left\{
\begin{array}{lll}
w_1 & = &
Q_{p} \left(\left[2(\lambda_{p} -6)z(z-1) -2\right]\alpha (z)
+\frac{4}{7}\left[(\lambda_{p} +6)z(z - \frac{1}{2})(z-1)\right]
\beta (z)\right)\\
w_2 & = &
Q_{p}\left( 2(1-z)\left[(\lambda_{p} -6)z +3\right]\alpha (z)
-\frac{4}{7}\left[(\lambda_{p} +6)z(z - 1)(z-\frac{3}{2})\right]
\beta (z)\right)\\
w_3 & = &
Q_{p} \left( \left[-2(\lambda_{p} -6)z(z-1) +3\right]\alpha (z)
-\frac{4}{7}\left[(\lambda_{p} +6)z(z - \frac{1}{2})(z-1)\right]
\beta (z) \right)
\end{array}
\right.
\ea
with
$\lambda_{p} =(p^2 +3)$.

The above conditions leave the overall normalisation 
of the bitensor undetermined.
To fix the normalisation constant $Q_{p}$ we contract the indices in the
coincident limit $z \rightarrow 1$. This yields \cite{Allen}
\begin{equation}
W^{(p)\ ij}_{ij}(\Omega,\Omega)= \sum_{{\cal P}lm}q_{ij}^{(p){\cal P}lm}
(\Omega) q^{(p){\cal P}lm\ ij}(\Omega)^{*}
 =30Q_{p}\alpha (1).
\end{equation}
By integrating over the three-sphere and using the normalisation condition
(\ref{norm}) on the tensor harmonics one obtains
$Q_{p} = -\frac{p^2 +4}{30 \pi^2 \alpha(1)}$.

Notice that (\ref{hyper}) is precisely the hypergeometric 
differential equation, which has a pair of 
independent solutions $\alpha(z)=\  _2F_{1}(3+ip,3-ip,7/2,z)$ and \newline
$\alpha(1-z) =\  _2F_{1}(3+ip,3-ip,7/2,1-z)$.
The former of these solutions is singular at $z=1$, i.e. for coincident points
on the three-sphere, and the latter is singular for opposite points.
The solution for $\beta (z)$ follows from (\ref{bet}) and is given by
\begin{equation}
\beta(z) =\  _2F_{1}(4-ip,4+ip,9/2,z).
\end{equation}
The hypergeometric functions are related by the transformation formula 
(eq.[15.3.6] in \cite{Abram})
\begin{center}
\begin{eqnarray}\label{relat}
_2F_{1}(a,b,c,z)
=\frac{\Gamma (c)\Gamma (c-a-b)}{\Gamma (c-a) \Gamma (c-b)}
\ _2F_{1}(a,b,a+b-c,1-z)&&\nonumber\\
 +
\frac{\Gamma (c)\Gamma (a+b-c)}{\Gamma(a)\Gamma(b)}(1-z)^{c-a-b}
\ _2F_{1}(c-a,c-b,c-a-b,1-z).&&
\end{eqnarray}\end{center}
Only for the eigenvalues of the Laplacian on $S^3$, i.e.
$p=in\ (n\geq 3)$,
the term on the second line vanishes for $_2F_{1}(3+ip,3-ip,
7/2,z)$. 
For these special values, $\alpha (z)$ and $\alpha (1-z)$ are no longer 
linearly independent but related by a factor
of $(-1)^{n+1}$, and they are both regular
for any angle on the three-sphere. In fact, the hypergeometric series
terminates for these parameter values and the hypergeometric functions
reduce to Gegenbauer polynomials $C^{(3)}_{n-3}(1-2z )$.
We have a choice between using 
$\alpha (z)$ and $\alpha (1-z)$ in the bitensor for these values of
$p$. Since $F(1-z) \rightarrow 1$
for coincident points, it is more natural to choose 
$\alpha(1-z)$ in the bitensor
appearing in the Euclidean Green function (\ref{disc}).
However, to obtain the Lorentzian correlator, we had to express
the discrete sum (\ref{disc}) as a contour integral. Since the Euclidean 
correlator obeys a differential equation with 
a delta function source at $\mu =0$, we must
maintain regularity of the integrand at $\mu = \pi$ when extending the 
bitensor in the complex $p$-plane. In other words, for generic $p$,
we need to work with the solution $\alpha (z)$, rather than $\alpha (1-z)$.
Therefore, in order to write the Euclidean correlator
as a contour integral, we first
have replaced $F(1-z)$ by $F(z)(-1)^{n+1}$, by applying
(\ref{relat}) to (\ref{disc}), and we then have continued the
latter term to $- (\cosh p\pi)^{-1}\ _2F_{1}(3+ip,3-ip,{7\over 2} ,z)$.

We conclude that
the properties of the bitensor appearing in the tensor correlator
completely determine its form.
Notice that in terms of the label $p$ we have obtained a
'unified' functional description of the bitensor
$W^{iji'j'}_{(p)}$ on $S^3$ and $H^3$. Its explicit form
is very different in both cases however, because the label $p$ takes on
different values. But it is precisely this description
that has enabled us in
Section IV to analytically
continue the correlator from the Euclidean instanton
into open de Sitter space without Fourier decomposing it.
We shall conclude this Appendix by giving the explicit formulae for
the coefficient functions of the bitensor $W^{L(p)}_{iji'j'}$ appearing
in our final result (\ref{lorcorf}). With this description, they can be
obtained by analytic continuation from $S^3$.

To perform the continuation to $H^3$ 
we note that the geodesic separation $\mu$ on $S^3$
continues to $-i\chi$ where $\chi$ is the comoving separation on $H^3$.
Hence 
the hypergeometric functions on $H^3$ are defined by analytic continuation
(eq. 15.3.7 in \cite{Abram}) and
may be expressed
in terms of associated Legendre 
functions as
\ba \label{app3a}
\left\{
\begin{array}{ll}
\alpha(z) & =15\sqrt{\frac{\pi}{2}}(-\sinh \chi
)^{-5/2} P^{-5/2}_{-1/2 +ip}(-\cosh \chi),
\\
\beta(z) &=15\sqrt{\frac{\pi}{2}}(-\sinh \chi
)^{-7/2} P^{-7/2}_{-1/2 +ip}(-\cosh \chi).
\end{array}
\right.
\ea
Using the relation
$-\cosh (\chi) = \cosh (\chi -i\pi)$, 
the Legendre functions on $H^3$ may be expressed as
\ba \label{app3}
\left\{
\begin{array}{lll}
P^{-5/2}_{-1/2 +ip}(-\cosh \chi) & = &
\sqrt{\frac{2}{-\pi \sinh \chi}}(1+p^2)^{-1}(4+p^2)^{-1}\left[
-3 \coth \chi \cosh p(\pi +i\chi)\right.\\
& & \left. -\frac{i\sinh p(i\chi +\pi)}{2p}\left( (2-p^2)
(1+\coth^2 \chi) +(4+p^2)\mathrm{cosech}^2 \chi \right) \right]\\
P^{-7/2}_{-1/2 +ip}(-\cosh \chi)& = &
\sqrt{\frac{2}{-\pi \sinh \chi}}(1+p^2)^{-1}(4+p^2)^{-1}(9+p^2)^{-1}\times\\
& & \left[
\cosh p(\pi +i\chi)(p^2  -11 - 15 \mathrm{cosech}^2 \chi)\right.\\
& & \left.
-6\frac{i\sinh p(i\chi +\pi)}{p}\left( (1-p^2)\coth^3 \chi
+(p^2+\frac{3}{2})\coth \chi \ \mathrm{cosech}^2 \chi \right) \right]
\end{array}
\right.
\ea
In the text, we have extracted the factors $e^{\pm p\pi}$ in these expressions 
in order to make contact with the usual description of the tensor correlator
in terms of tensor harmonics on $H^3$.
The coefficient functions of the bitensor 
$W_{\ iji'j'}^{L (p)}(\chi)$
in our final result (\ref{lorcorf})
for the tensor correlator are
\ba \label{app4}
\left\{
\begin{array}{lll}
w_1 & = &
\frac{\mathrm{cosech^5 \chi}}{4\pi^2 (p^2 +1)}
\left[\frac{\sin p\chi}{p}(3+(p^2 +4)\sinh^2\chi - 
p^2 (p^2 +1) \sinh^4 \chi)
\right.\\
& & \ \ \ \ \ \ \ \ \ \left. 
-\cos p\chi (3/2 + (p^2 +1) \sinh^2 \chi )\sinh 2\chi\right]\\
w_2 & = &
\frac{\mathrm{cosech^5 \chi}}{4\pi^2 (p^2 +1)}
\left[\frac{\sin p\chi}{p}(3+12\cosh \chi -3p^2(1+2\cosh \chi )
\sinh^2\chi\right.\\
& & \ \ \ \ \ \ \ \ \ \left. + p^2 (p^2 +1) \sinh^4 \chi) 
+\cos p\chi (-12-3 \cosh \chi \right.\\
& & \ \ \ \ \ \ \ \ \ \left.+2(p^2 -2) \sinh^2 \chi +2(p^2 +1)\cosh \chi 
\sinh^2 \chi) \sinh \chi\right]\\
w_3 & = &
\frac{\mathrm{cosech^5 \chi}}{4\pi^2 (p^2 +1)}
\left[\frac{\sin p\chi}{p}(3 -3p^2\sinh^2\chi +p^2 (p^2 +1) \sinh^4 \chi)\right.\\
& & \ \ \ \ \ \ \ \ \ \left. 
+\cos p\chi (-3/2 + (p^2 +1) \sinh^2 \chi )\sinh 2\chi\right]\\
\end{array}
\right.
\ea
As mentioned before, the bitensor $W^{L(p)}_{iji'j'}$ equals the sum
(\ref{bit}) of the rank-two tensor eigenmodes with eigenvalue 
$\lambda_{p}=-(p^2 +3)$ of the Laplacian on $H^3$.
For $\chi \rightarrow 0$ these functions converge and
they exponentially decay at large geodesic distances.


\begin{thebibliography}{99}

\bibitem{Hertog}T.Hertog and N.Turok, submitted to 
Phys. Rev. D, astro-ph/9903075.
%\bibitem{Garriga}J.Garriga, X.Montes, M.Sasaki and T.Tanaka,
%"Spectrum of Cosmological Perturbations in the One-Bubble Open Universe",
%astro-ph/9811257 (1998).
\bibitem{allencald}B.Allen and R.Caldwell, unpublished manuscript (1996).
\bibitem{Tanaka}T.Tanaka and M.Sasaki,
% "The Spectrum of Gravitational Wave 
%Perturbations in the One-Bubble Open Inflationary Universe", 
Prog. Theor. Phys.
$\mathbf{97}$, 243 (1997).
\bibitem{Mottola}I.Antionadis and E.Mottola,
% "Graviton Fluctuations in de Sitter Space",
J. Math. Phys. $\mathbf{32}$, 1037 (1991).
\bibitem{Hartle}J.B.Hartle and S.W.Hawking,
% "The Wave Function of the Universe",
Phys. Rev. $\mathbf{D28}$, 2960 (1983).
\bibitem{Hawking}S.W.Hawking and N. Turok,
% "Open Inflation without False Vacua",
Phys. Lett. $\mathbf{B425}$, 25 (1998).
\bibitem{Bucher}M.Bucher, A.S.Goldhaber and N.Turok,
% "An Open Universe from Inflation", 
Phys. Rev. $\mathbf{D52}$, 3305 (1996).
\bibitem{Gratton}S.Gratton and N.Turok, Phys. Rev.
$\mathbf{D60}$, 123507 (1999), astro-ph/9902265.
\bibitem{Coleman}S.Coleman and F.De Luccia,
% "Gravitational Effects on and of Vacuum Decay", 
Phys. Rev. $\mathbf{D21}$, 3305 (1980).
%\bibi{Gar} J.Garriga, hep-th/9803210 (1998).
%\bibitem{Cohn}M.Bucher and J.D.Cohn, 
%"Primordial Gravitational Waves from Open Inflation", 
%Phys. Rev. $\mathbf{D55}$, 7461 (1997).
%\bibitem{sasrefs} Yamamoto, M. Sasaki and T. 
%Tanaka, Ap. J. {\bf 455}, 412 (1995);
%Phys. Rev. {\bf D54}, 5031 (1995); T. Tanaka and M. Sasaki, Phys. Rev. {\bf 
%D59} (1999) 023506; see also the discussion of the path integral
%approach in T. Tanaka and M. Sasaki, Prog. Theor. Phys. {\bf 88} (1992) 503.
\bibitem{Stewart}J.M.Stewart,
% "Perturbations of Friedmann-Robertson-Walker cosmological models",
Class. Quantum Grav. $\mathbf{7}$, 1169 (1990).
\bibitem{Bruce}B.Allen,
% "Vacuum States in de Sitter Space",
Phys. Rev. $\mathbf{D32}$, 3136 (1985). 
%\bibitem{Sasaki}M.Sasaki, T.Tanaka, Y.Yakushige, 
%"Wall Fluctuation Modes and Tensor CMB Anisotropy in Open Inflation Models",
%Phys. Rev. $\mathbf{D56}$, 616 (1997).
%\bibitem{xavi} J. Garriga, X. Montes, M. Sasaki and T. Tanaka, Nucl. Phys
%{\bf B513}, 343 (1998).
%\bibitem{LST}  A. Linde, M. Sasaki, T. Tanaka, astro-ph/9901135 (1999).
%\bibitem{Kodama}H.Kodama and M.Sasaki, Prog. Theor. Phys. Suppl.
%$\mathbf{78}$, 1 (1984).
\bibitem{Mukhanov}V.F.Mukhanov, H.A.Feldman and R.H.Brandenberger,
% "Theory of Cosmological Perturbtions", 
Phys. Rep. $\mathbf{215}$, 203 (1992).
%\bibi{newton} R.G. Newton, {\it Scattering Theory of Waves and Particles},
%McGraw Hill Book Co. (1966), p. 334.
\bibitem{Erdelyi}A.Erdelyi, \emph{Tables of Integral Transforms}, 
Mc.Graw-Hill Book Company (1954).
%\bibitem{garrigadisc} We thank J. Garriga for a 
%discussion of this point. 
%\bibitem{Garr}J.Garriga, 
%"Bubble Fluctuations in $\Omega <1$ Inflation",
%Phys. Rev. $\mathbf{D54}$, 4764 (1996).
%\bibitem{Bellido}J.Garcia-Bellido, 
%"Metric Perturbations from Quantum Tunneling in Open Inflation", 
%Phys. Rev. $\mathbf{D54}$, 2473 (1996).
\bibitem{Sachs}R.Sachs and A.Wolfe, 
%"Perturbations of a Cosmological Model and Angular Variation of the Microwave %Background", 
Ap. J. $\mathbf{147}$, 73 (1967).
\bibitem{gratthertog}S.Gratton, T.Hertog and N.Turok, 
DAMTP preprint, astro-ph/9907212.
\bibitem{reall}S.W.Hawking, T.Hertog and H.S.Reall, in preparation (2000).
%\bibitem{Tomita}K.Tomita, 
%"Tensor Spherical and Pseudo-Speherical Harmonics in 
%Four-Dimensional Spaces", 
%Prog. Theor. Phys. $\mathbf{68}$, 310 (1982).      
%\bibitem{Garcia}D.H.Lyth, D.Wands, J.Garcia-Bellido and A.Liddle, 
%"Normalisation of Modes in an Open Universe", 
%Phys. Rev. $\mathbf{D55}$, 4596 (1997).
\bibitem{Jacob}B.Allen and T.Jacobson,
% "Vector Two-Point Functions in Maximally Symmetric Spaces",
Commun. Math. Phys. $\mathbf{103}$, 669 (1986).
\bibitem{Allen}B.Allen,
% "Maximally Symmetric Spin-Two Bitensors on $S^3$ and $H^3$", 
Phys. Rev. $\mathbf{D51}$, 5491 (1995).
\bibitem{Abram}M.Abramowitz and I.Stegun (Eds.), \emph{Handbook of
Mathematical Functions}, National Bureau of Standards (1972).


\end{thebibliography}
\end{document}